\documentclass[useAMS,referee,usenatbib, subcaption]{biom}
\def\bSig\mathbf{\Sigma}

\usepackage[export]{adjustbox}
\usepackage{float,booktabs,multirow,subcaption, hyperref}
\usepackage{rotating}
\usepackage{amssymb,amsmath}

\title[Supervised Robust Profile Clustering]{Supervised Robust Profile Clustering}

\author{Briana Stephenson$^{1,*}$\email{bstephenson@hsph.harvard.edu}, 
Amy Herring$^{2,3,4}$, and Andrew Olshan$^{5}$ \\
$^{1}$Department of Biostatistics, Harvard University, Boston, Massachusetts, U.S.A. \\
$^{2}$Department of Statistical  Science, Duke University, Durham, North Carolina, U.S.A.\\
$^{3}$Duke Global Health Institute, Duke University, Durham, North Carolina, U.S.A.\\
$^{4}$Department of Biostatistics and Bioinformatics, Duke University, Durham, North Carolina, U.S.A.\\
$^{5}$Department of Epidemiology, University of North Carolina, Chapel Hill, North Carolina, U.S.A.\\
}

\begin{document}





\pagerange{\pageref{firstpage}--\pageref{lastpage}} 




\label{firstpage}


\begin{abstract}
In many studies, dimension reduction methods are used to profile participant characteristics. For example, nutrition epidemiologists often use latent class models to characterize dietary patterns. One challenge with such approaches is understanding subtle variations in patterns across subpopulations. Robust Profile Clustering (RPC) provides a dual flexible clustering model, where participants may cluster at two levels: (1) globally, where participants are clustered according to behaviors shared across an overall population, and (2) locally, where individual behaviors can deviate and cluster in subpopulations. We link clusters to a health outcome using a joint model. This model is used to derive dietary patterns in the United States and evaluate case proportion of orofacial clefts. Using dietary consumption data from the 1997-2009 National Birth Defects Prevention Study, a population-based case-control study, we determine how maternal dietary profiles are associated with an orofacial cleft among offspring. Results indicated that mothers who consumed a high proportion of fruits and vegetables compared to meats, such as chicken and beef, had lower odds delivering a child with an orofacial cleft defect. 
\end{abstract}

%


\maketitle


%

\section{Introduction}
\label{s:intro}

Robust profile clustering (RPC) is a statistical framework that is able to discriminate features of a diverse population that contribute to the general population as opposed to a subpopulation \citep{stephenson2019robust}. RPC provides a multilevel approach that allows subjects and variables to cluster both locally, within a subject's respective subpopulation, and globally, across all subpopulations. 

This clustering approach identifies features that are likely to describe differences in global behaviors compared to those relevant to specific subpopulations. Linking these derived clusters to an outcome requires a  flexible regression model. In the case of RPC, subjects can have both global and local cluster assignments that differ from one variable to another. Global clusters include shared behaviors across the overall population. Local clusters highlight a subset of behaviors shared amongst subjects belonging to a specific subpopulation. Thus two subjects from different subpopulations could be in the same global cluster but differ slightly in behavioral patterns. Similarly, two subjects from the same subpopulation could belong to the same local cluster, sharing behaviors amongst a select set of variables, but belong to different global clusters, where the remaining set of variables reflect different behaviors across the overall population. For example, suppose global clusters were used to describe overall dietary behaviors across the United States, and local clusters were used to describe dietary behaviors within an individual state (subpopulation). If overall dietary patterns indicated a general non-consumption of oranges, but residents in Florida indicated a very high intake of oranges, then oranges would be considered a locally deviated food that would have a separate consumption pattern for Florida residents only, without changing estimates of national dietary patterns.
 
Of interest to researchers is how a complex exposure, such as maternal diet, may influence health outcomes. We approach this question by developing a supervised clustering framework. This allows subjects who are more likely to exhibit a health outcome to cluster in accordance with their behavior profile. 

We organize this paper as follows. Section 2 introduces the supervised RPC joint model. Section 3 describes the algorithm for posterior computation and inference. Section 4 compares the utility of the model with a traditional latent class model via a simulation study. Section 5 presents an application of the model using the NBDPS data to explore associations between maternal diet and oral cleft birth defects. We conclude with a short discussion on extensions and further methodological developments in Section 6.

\section{Model}
\label{s:model}
\subsection{Background}
Latent class models are a common technique used to identify subgroups within a heterogeneous population that share a common set of behaviors or characteristics \citep{vermunt2003latent}. This can extend to a supervised or joint latent class model when it is believed that these subgroups share the same probability of an event or outcome \citep{proust2014joint,larsen2004joint,bigelow2009bayesian}. The membership assignment of the derived latent class is treated as a covariate in a regression model (e.g. logistic, probit, Cox proportional hazards) \citep{lin2002latent,moustaki2005latent,ferrat2016four,sotres2013maternal}. 
A limitation to the standard latent class model is that subjects assigned to the same cluster are assumed to share the same responses to all variables included in the model. This is unrealistic when handling heterogeneous populations that are composed of different subpopulations that are likely to deviate for certain variables included in the model. This is further exacerbated in high-dimensional cases, where there are a large number of variables. Subpopulation effects can mask the contribution of some variables and create clusters that are less informative to questions of interest. Potentially masked effects can be avoided by removing a subset of variables found to be noninformative from the model \textit{a priori} or jointly with estimation of the supervised model. Only the subset of variables identified as important are included in the final model. These variables are selected based on the predictive power or correlation to the outcome \citep{bair2004semi,houseman2006feature,desantis2012supervised}. However, these methods also do not account for subpopulation differences, where the degree of importance can change depending on the variable. A flexible model is needed to allow the number of variables important in contributing to the characteristics of the overall population to vary from subpopulation to subpopulation, while also allocating those subpopulation-specific variables to a separate local clustering model.

\subsection{Robust Profile Clustering Predictor Model}
The RPC process utilizes both subpopulation-specific and general population information to generate global and local cluster assignments for each subject based on their observed responses to a set of exposure variables. It is induced through a 3-component probabilistic model: (\textit{i}) the global clustering, $C_i$, (\textit{ii}) the local clustering membership $L_{ij}$, and (\textit{iii}) the global deviation indicators $G_{ij}$. Let $n$ be the number of subjects in the study population that is comprised of $S$ subpopulations. Each subpopulation $s \in (1,\ldots,S)$ contains $n_s$ subjects. Let $x_i =(x_{i1},\ldots,x_{ip})$ denote the observed data vector for subject $i$ for all predictor variables. 

We select a Bayesian framework to stabilize parameter estimation of the model and allow structural flexibility with limited information \textit{a priori} of the data. Similar to the latent class model first introduced by \citet{lazarsfeld1968latent}, the global clustering component is generated from a finite mixture model. Assuming no prior knowledge of the size of these clusters, a flat symmetric Dirichlet distribution is used to estimate the weights of these clustering components, such that $Pr(C_i=h)=\pi_h, \pi =(\pi_1,\ldots, \pi_{K_0}) \sim Dir(\alpha_0)$, where $\alpha_0$ is a fixed Dirichlet hyperparameter selected for the overall population. The local clustering component may also be generated from a finite mixture model, for each subpopulation $s$ and variable $j \in (1,\ldots,p)$, where $Pr(L_{ij}=l|s_i=s)=\lambda_l^{(s)}, \lambda^{(s)} = (\lambda_1^{(s)}, \ldots, \lambda_{K_s}^{(s)}) \sim Dir(\alpha_s)$, where $\alpha_s$ is a fixed Dirichlet hyperparameter selected for subpopulation $s$.  If the number of clusters is unknown, in either case, a conservative upper bound that far exceeds the expected number of clusters in the model can be used to mimic an overfitted finite mixture model \citep{rousseau2011asymptotic}.

Each variable is assumed to either contribute to the overall population (at a global level) or or to the known subpopulation of the subject (at the local level). The global/local allocation component, $G_{ij}$, is drawn from a Beta-Bernoulli process, where $G_{ij}|s_i=s \sim Bern(\nu_j^{(s)})$, $\nu_j^{(s)} \sim \mathcal{B}(1,\beta^{(s)})$, for each variable $j \in (1,\ldots, p)$, and $G_{i\cdot}$ is the vector variable containing all $G_{ij}$. Applying this model to dietary pattern analysis in the United States, $s$ would index each state included in the dataset, and $j$ would index a specific food item from a list of $p$ foods. The set of observed foods by individual $i$ is described as $x_i = (x_{i1},\ldots,x_{ip})$. 

The set of global cluster parameters, $\Theta_{0\cdot\cdot} = \{\theta_{0jC_i}\}_{j=1}^p$ describe overall (national) population behaviors, such that $\theta_{0jh,r}$ is the probability of consumption at the global level of food item $j$ assigned to global profile $h$ at the $r \in (1,\ldots, d_j)$ consumption level. The set of local cluster parameters, $\Theta_{1\cdot\cdot}^{(s)} = \{\theta_{1jL_{ij}}^{(s)}\}_{j=1}^p$, describe local behaviors specific to subpopulation $s$, such that $\theta_{1jl,r}^{(s)}$ is the probability of consumption at the local level of food item $j$ assigned to local profile $l$ within subpopulation $s$ at the $r \in (1,\ldots, d_j)$ consumption level. The maximum number of levels for each variable $j$ is denoted $d_j$. The density of these parameters corresponds to the data type of the observed variables. We focus this paper on multinomial categorical data, but more generally can accommodate other measurement scales (e.g. continuous, count). The induced subject-specific likelihood illustrated in (\ref{rpclikeli}) demonstrates that variable contributions are split based on whether they are allocated to the global model, when $G_{ij}=1$, or the subpopulation/local model, when $G_{ij}=0$. 
\begin{equation}\label{rpclikeli}
f(\mathbf{x}_i|\mathbf{\pi}, \mathbf{\lambda}^{(s_i)},G_{i\cdot},\Theta_{0\cdot\cdot}, \Theta_{1\cdot\cdot}^{(s_i)}) = \left[\sum_{h=1}^{K_0} \pi_h \prod_{j:G_{ij}=1}^p\prod_{r=1}^{d_j} \theta_{0jh,r}^{\mathbf{1}(x_{ij}=r)}\right] \prod_{j:G_{ij}=0}^p\left[\sum_{l=1}^{K_s} \lambda_{l}^{(s_i)} \prod_{r=1}^{d_j} (\theta_{1jl,r}^{(s_i)})^{\mathbf{1}(x_{ij}=r)}\right]
\end{equation}

Partitioning out local deviations into a separate clustering structure by subpopulation enables a more parsimonious set of global clusters compared to a traditional latent class model. Due to the restriction of a single global clustering structure, more clusters will form to accommodate seemingly minor differences that may exist between them. The RPC reduces this abundance by determining if those minor differences are specific to a subpopulation and allowing them to be redacted from the global clustering model and redirected to a local clustering model. 

\subsection{Probit Regression Response Model}
We wish to evaluate the association between global cluster membership and an outcome, $y_i$. In our case, $y_i=1$ indicates an orofacial cleft. We represent the set of subject-specific observed covariates as $W_i$. These covariates include study site (state) location of the subject, designated by $s$, global profile membership, $C_i$, and a collective set of $q$ demographic variables (e.g. age, education, race, etc.). The fixed effect for  each of these covariates is estimated by $\xi_h \in (1,\ldots, S+K_0+q)$.

\begin{equation}\label{eq:probit}
P(y_i =1|W_i, C_i, s_i, \xi) = \Phi \bigg(  \sum_{h=1}^{S+K_0+q-1} W_i \xi_{h} \bigg)
\end{equation}

Using the global cluster assignments generated from the predictor model, the probit response model in (\ref{eq:probit}) is constructed, leading to the joint likelihood model. 

\begin{equation}\label{eq:jointlikeli}
\mathcal{L}_i = \left[\sum_{h=1}^{K_0} \pi_h \prod_{j:G_{ij}=1}^p\prod_{r=1}^{d_j} \theta_{0jh,r}^{\mathbf{1}(x_{ij}=r)} \Phi(W_i\xi)^{y_i}[1-\Phi(W_i\xi)]^{1-y_i} \right] \prod_{j:G_{ij}=0} \left[\sum_{l=1}^{K_s} \lambda_{l}^{(s_i)} \prod_{r=1}^{d_j} (\theta_{1jl,r}^{(s_i)})^{\mathbf{1}(x_{ij}=r)}\right]
\end{equation}

To preserve re-ordering from label switching within the probit model, a cell means coding structure was implemented. To allow parameter estimation of the response model, we implement the sampling approach of a latent variable $Z_i$, where $Y_i = \mathbf{1}(Z_i > 0)$, and $Z_i$ follows a linear regression, as introduced by \citet{albert1993bayesian}.

\section{Methods}
\subsection{Parameter Estimation}\label{parmest}
The joint model is intended to estimate both parameters from the RPC-predictor model and the probit response model concurrently in an MCMC algorithm. When the number of global clusters is not known {\it a priori}, an overfitted finite mixture model is used to determine the appropriate number of global clusters to fit in the joint model. While superior to many approaches, overfitted mixture models are still prone to generate both extraneous and redundant clustering \citep{van2015overfitting}. A similarity matrix was calculated based on the MCMC output of the RPC model containing pairwise posterior probabilities of two subjects being clustered together in a given iteration.  Hierarchical clustering was then performed on the similarity matrix, using the complete linkage approach \citep{krebs1989ecological,medvedovic2002bayesian}. The appropriate number of clusters was then determined from the hierarchical clustering dendrogram, if needed, and applied to the dataset for a single cluster estimate, based on the MCMC output. The supervised RPC joint model was then applied using the appropriate number of global clusters. Posterior median estimates of all parameters were relabelled and calculated.

\subsection{MCMC Algorithm}
While informative priors may be used, we select a set of reasonable but minimally informative priors as a default. We set up the model with $\text{Ga}(1,1)$ prior on the Beta-Bernoulli process hyperparameter $\{\beta^{(s)}\}_{s=1}^S$. For our multinomial categorical predictors where all variables contain $d$ different response levels, we select a flat symmetrical Dirichlet prior $(\eta=1)$, for $\theta_{0jh,\cdot} = \{\theta_{0jh,1},\ldots,\theta_{0jh,d}\}$ for all $j\in (1,\ldots,p)$ variables, in each global cluster $h \in (1,\ldots,K)$, and likewise for the local cluster density parameters. The covariates included in the regression model to describe a subject's demographic information $W_i$ and global profile membership $C_i$ are binary or scaled (normalized) variables. These variables are treated as fixed effects and fit with a multivariate normal prior, such that $\xi \sim MVN(\mu_0,\Sigma_0)$, where $\mu_0$ is a $q-$length vector and each element is drawn from a standard normal distribution. $\Sigma_0$ is a diagonal matrix, where each entry is drawn from an Inverse Gamma distribution with shape, $a_\sigma = \frac{5}{2}$ and scale, $b_\sigma=\frac{5}{2}$.  To encourage sparsity, the Dirichlet hyperparameter of the global and local finite mixtures was selected as $\alpha=\frac{1}{K}$ \citep{rousseau2011asymptotic}. The latent response variable, $z_i$, was drawn from a truncated normal distribution, based on if the subject was designated as an oral cleft defect case $(y_i = 1)$ or healthy control $(y_i = 0)$, as detailed below. 

\textbf{Data Model: } 
\begin{equation} \label{eq:latprobit}
\begin{aligned}
P(y_i=1|s_i, \xi, C_i,W_i) &= \Phi (W_i \xi) \\
Z_i &= \Phi(\xi_1 + \sum_{s=2}^{S} \mathbf{1}(s_i=s) \xi_{s} + \sum_{k=2}^{K_0} \mathbf{1}(C_i =k)\xi_{S+k-1} + W_i\xi_{dem} + \epsilon_i )\\
& \mbox{where } \epsilon_i \sim N(0,1) \\
Z_i &= \begin{cases}
	> 0, \mbox{ when } y_i = 1 \\
	\le 0, \mbox{ when } y_i = 0
\end{cases}
\end{aligned}
\end{equation}

Posterior probabilities were calculated to determine if a parameter was likely to contribute to a higher or lower proportion of orofacial cleft case. This was calculated such that $Pr(\xi_l > 0) = \frac{1}{M} \sum_{t=BURN+1}^M \mathbf{1}(\xi_l^{(t)} > 0)$, where M is the total number of MCMC iterations after burn-in, and $\xi_l^{(t)}$ is the probit coefficient $l$ estimate at iteration $t$. 

\subsubsection{Posterior Computation and Inference}\label{gibbs}
We propose a simple Gibbs sampler for posterior computation.

\begin{enumerate}
\item Update the global component indicators $(G_{ij} \mid s_i=s) \sim \text{Bern}(p_{ij})$, where $$p_{ij}=\frac{\nu_j^{(s)}\prod_{r=1}^{d_j} \Theta_{0jC_i,r}^{\mathbf{1}(x_{ij}=r)}}{\nu_j^{(s)}\prod_{r=1}^{d_j}\Theta_{0jC_i,r}^{\mathbf{1}(x_{ij}=r)}+(1-\nu_j^{(s)})\prod_{r=1}^{d_j} (\Theta^{(s)}_{1jL_{ij},r})^{\mathbf{1}(x_{ij}=r)}}$$ for each subject $i \in (1,\ldots,n)$ with respective subpopulation index $s$. 

\item Update global cluster index $C_i$, $i=1,\ldots,n$ from its multinomial distribution where $$Pr(C_i=h) = \frac{\pi_h \prod_{j} \prod_{r=1}^{d_j} \Theta_{0jh,r}^{\mathbf{1}(x_{ij}=r, G_{ij}=1)}\Phi(W_i\xi)^{y_i}[1-\Phi(W_i\xi)]^{1-y_i}}{\sum_{l=1}^K \pi_l \prod_{j} \prod_{r=1}^{d_j} \theta_{0jl,r}^{\mathbf{1}(x_{ij}=r, G_{ij}=1)}\Phi(W_i\xi)^{y_i}[1-\Phi(W_i\xi)]^{1-y_i}}.$$

\item Update local cluster index $L_{ij}$ for all $i:s_i=s$ and $j=1,\ldots,p$, repeating for each $s$, from its multinomial distribution conditional on $s_i=s$ where $$Pr(L_{ij}=h)=\frac{\lambda^{(s)}_h \prod_{r=1}^{d_j} \left(\theta^{(s)}_{1jh,r}\right)^{\mathbf{1}(x_{ij}=r, G_{ij}=0)}}{\sum_{l=1}^K \lambda^{(s)}_l \prod_{r=1}^{d_j} \left(\theta^{(s)}_{1jl,r}\right)^{\mathbf{1}(x_{ij}=r, G_{ij}=0)}}.$$

\item Update the global clustering weights $$\pi=(\pi_1,\ldots,\pi_K) \sim \text{Dir}\left(\alpha_0+\sum_{i=1}^n \mathbf{1}(C_i=1),\ldots, \alpha_0+\sum_{i=1}^n \mathbf{1}(C_i=K)\right).$$

\item Update the local clustering weights in subpopulation $s$, $$\lambda^{(s)}=\left(\lambda_1^{(s)},\ldots,\lambda_K^{(s)}\right) \sim \text{Dir}\left(\alpha_s + \sum_{i:s_i=s} \sum_{j=1}^p \mathbf{1}(L_{ij}=1), \ldots, \alpha_s + \sum_{i:s_i=s} \sum_{j=1}^p \mathbf{1}(L_{ij}=K)\right).$$

\item Update the multinomial parameters, where $\sum_{i:G_{ij}=1,C_i=h} \mathbf{1}(x_{ij}=r)$ is the total number of subjects allocated to global cluster $h$ for variable $j$ with an observed consumption level of $r$, letting $\eta=1$. Similarly, $\sum_{i:G_{ij}=0,L_{ij}=h,s_i=s} \mathbf{1}(x_{ij}=r)$ is the total number of subjects allocated to local cluster $h$ within subpopulation $s$ for variable $j$ and an observed consumption level of $r$.
\begin{align*}
\theta_{0jh,\cdot} &\sim \text{Dir}\left(\eta+\sum_{i:G_{ij}=1,C_i=h} \mathbf{1}(x_{ij}=1),\ldots,\eta+\sum_{i:G_{ij}=1,C_i=h} \mathbf{1}(x_{ij}=d)\right) \\
\theta^{(s)}_{1jh,\cdot} &\sim \text{Dir}\left(\eta+\sum_{i:G_{ij}=0,L_{ij}=h,s_i=s} \mathbf{1}(x_{ij}=1),\ldots,\eta+\sum_{i:G_{ij}=0,L_{ij}=h,s_i=s} \mathbf{1}(x_{ij}=d)\right)
\end{align*}

\item Update $\nu_j^{(s)} \sim \text{Be}(1+\sum_{i:s_i=s} G_{ij}, \beta^{(s)} + \sum_{i:s_i=s} (1-G_{ij}))$.

\item Update Beta-Bernoulli hyperparameter: $\beta^{(s)} \sim \text{Ga}(a_{\beta}+p, b_{\beta} - \sum_{j=1}^p \log(1-\nu_j^{(s)}))$.

\item Update regression coefficients: $\xi \sim MVN((\Sigma_0^{-1} + W'W)^{-1}(\Sigma_0^{-1}\xi_0 + W'Z) ,(\Sigma_0^{-1} + W'W)^{-1})$, where $\xi_0 \sim MVN(\mu_0, \Sigma_0)$.

\item Update latent response variable: $Z_i|W_i,\xi, y_i \sim \begin{cases}
	N_{z_{i} \le 0}(W_i\xi,1) \mbox{ when } y_i = 0  \\
	N_{z_{i} > 0}(W_i\xi,1) \mbox{ when } y_i = 1
	\end{cases}$
\item Calculate probit likelihood,$\Phi(W_i\xi)^{y_i}[1-\Phi(W_i\xi)]^{1-y_i}$, for all $K$ clusters and $i \in (1,\ldots, n)$ subjects. 
\end{enumerate}

\section{Simulation}
We demonstrate the utility of our model using a simulation study, where we compare our model against the traditional latent class analysis (LCA) model. We generate our dataset by defining $S=4$ subpopulations containing $n_s=1200$ subjects each. Measurements are available of $p=50$ observed categorical variables that each contain $d_j=4$ response levels for all $j \in (1,\ldots, p)$ variables. Subjects were randomly assigned to one of three global profiles. Within each subpopulation, subjects were randomized into one of two local profiles. Global and local variables were preallocated based on a prefixed $\{\nu_j^{(s)}\}_{j=1}^p$, such that every simulated observation has at least some local deviation. 

Probability of an outcome was determined using a probit regression model with subpopulation and global cluster assignments introduced as covariates. We generate 500 datasets from the true model. The supervised RPC model was run using the Gibbs sampler algorithm outlined in Section \ref{gibbs}. Mean square error (MSE) was calculated to compare lower predicted proportion  $(\Phi(W_i\hat{\xi}))$ to true proportion $(\Phi(W_i\xi))$. Sensitivity was calculated by the proportion of subjects from a pre-assigned cluster that remained together in the cluster assigned by the supervised RPC model. 

The Deviance Information Criterion (DIC) was calculated for each model to evaluate model fitness of the supervised RPC joint model, compared to the traditional latent class model \citep{spiegelhalter2002bayesian}. To determine the appropriate number of classes in LCA suitable for these data, models were fit with an increasing number of classes until best fit was reached. The latent class model's log-likelihood function for the clustering predictor model is outlined in (\ref{eq:lca}), where $K$ denotes the number of clusters best fit in the model. 

 \begin{equation}\label{eq:lca}
 \log \mathcal{L}(x_i|\pi, \theta)= \sum_{i=1}^N \log \bigg(\sum_{h=1}^K \pi_h \prod_{j=1}^p \prod_{c=1}^{d_j} \theta_{jc|k}^{\mathbf{1}(x_{ij}=c)} \bigg) + y_i log(\Phi(W_i\xi)) + (1-y_i) log(1-\Phi(W_i\xi))
 \end{equation}

The deviance of the respective models were calculated using the log-likelihood of both the clustering predictor model (\ref{rpclikeli}) and the probit response model (\ref{eq:probit}). The DIC is calculated by taking the difference of the posterior mean log-likelihood derived at each iteration of the MCMC with the log-likelihood derived using the posterior mean estimates of the model parameters, denoted $\tilde{\theta}$ \citep{spiegelhalter2002bayesian}. Given the mixture of distributions used in both the supervised RPC and LCA, the calculation of DIC can be approximated as a missing data model \citep{celeux2006deviance}. Due to overfitting evident in DIC, several researchers have suggested an adaptation to the DIC that increases the penalty of complexity \citep{van2005dic,plummer2008penalized,ando2010predictive,van2012bayesian,spiegelhalter2014deviance}. We adapt this proposed penalization to $DIC_6$, such that
 
\begin{equation}\label{dic}
DIC_6^* = 3 \bar{D}(\theta) - 2 D(\bar{\theta}) = -6 \mathbf{E}_\theta [\log \mathcal{L}(y|\theta)|y] + 4 \log \mathcal{L}(y|\tilde{\theta})
\end{equation}

where $\bar{D}(\theta)$ denotes the posterior mean observed deviance and $D(\bar{\theta})$ the deviance of posterior mean. 

\subsection{Results} 

Across all 500 simulated sets, the supervised RPC successfully predicted a median of 3 global clusters, whereas the supervised LCA predicted an additional fourth global cluster. The supervised RPC had a considerable improvement of predicting the correct outcome compared to the supervised LCA $(MSE_{RPC} = 0.01, MSE_{LCA} = 0.07)$. This may be attributed to the additional cluster derived by the supervised LCA which led to a decreased goodness of fit $(DIC_{RPC} = 4.32\times10^5; DIC_{LCA}=4.64\times10^5)$ and decreased sensitivity $(RPC=100\%, LCA=87\%)$ in identifying the correct cluster assignment. 

Cluster patterns for each model were derived using the posterior modes calculated from the global cluster density parameters. Expected response for a given variable was defined as the response level containing the maximum posterior probability.  Web Figure 1 provides an illustration of the cluster patterns derived from both models using one of the 500 simulated datasets. 

The introduction of local deviations hybridized within the global profiles was pronounced in the modal patterns derived from the standard latent class model. Due to the limitation of a single global clustering assumption, the modal patterns derived in the LCA model introduced additional noise into the expected true cluster patterns. Ignoring noise, latent class 1 resembled the pattern in true cluster 3. Latent cluster 2 resembled a combination of the pattern expected in true cluster 1 and 2. Latent class 3 resembled the pattern expected in true cluster 2. Latent class 4 resembled the pattern expected in true cluster 1. The patterns from the supervised RPC model were easily identifiable to the true patterns with only one variable in global profiles 1 and 2 showing slight variation in results.  

In addition to an improved identification of the expected modal response, the supervised RPC model was also able to discriminate which variables are likely to be global or local $(MSE_\nu < 0.01)$. Variables with high probabilities were classified as global (shaded white in Web Figure 2). Variables with low probabilities were classified as local (shaded black in Web Figure 2).

\section{Application to National Birth Defects Prevention Study}
\subsection{NBDPS Dietary Data}
The National Birth Defects Prevention Study (NBDPS) is an observational, population-based, case-control study of birth defects in the United States \citep{yoon2001national}. Self-reported maternal dietary consumption data were collected by telephone interview from women at ten different centers across the United States from 1997-2009. Cases were defined as any live-born, stillborn, or electively terminated infant with an orofacial cleft defect.  Cleft defects were classified as cleft lip with cleft palate, cleft lip without cleft palate, or cleft palate alone. Controls were defined as any live-born infant without any birth defects and were randomly selected from birth certificates or hospital records. Exclusion criteria for this analysis included mothers with multiple births, a family history of clefts (33 controls, 224 cases), preexisting diabetes (109 controls, 86 cases), use of folate antagonist medications from the time period 3 months before pregnancy to the end of pregnancy (124 controls, 55 cases), extreme reported values in total caloric intake ($<500$ or $> 5000$ kcal) (314 controls, 125 cases), and missed more than 1 item in the food frequency questionnaire or data on folic acid/multi-vitamin supplement use  (227 controls, 73 cases), missing demographic covariates (61 controls, 21 cases). After exclusions, a total of 3409 cases and 8949 controls were included for analysis. 

\begin{table}[H]
\centering
\renewcommand{\arraystretch}{0.85}
\begin{tabular}{|l|l|c|c| }
\hline
\multicolumn{2}{ |c| }{ National Birth Defects Prevention Study} & Oral Cleft & Control \\
\multicolumn{2}{ |c| }{ Demographic Information } & (n=3409) & (n=8949) \\
\hline
Variable & Category & n (\%) & n (\%)\\ \hline
Race/Ethnicity & Non-Hispanic White & 2098 (62) & 5209 (58) \\
 & Non-Hispanic Black & 202 (6) &986 (11) \\
 & Hispanic & 848 (25) & 2112 (24) \\
 & Other & 261 (8) & 642 (7) \\ \hline
Age, in years & $< 25$ yrs & 1180 (33) & 2940 (35) \\
 & $\ge 25$ yrs & 2229 (67) & 6009 (65) \\
\hline
Education & HS or less & 1577 (46) & 3671 (41) \\
 & Beyond HS & 1832 (54) & 5278 (59) \\
\hline
BMI & Underweight, $BMI < 18.5$ & 200 (6) & 469 (5) \\
 & Normal, $18.5 \leq BMI < 25$ & 1706 (53) & 4647 (54) \\
 & Overweight/Obese, $BMI >25$ & 1338 (41) &  3458 (40) \\ 
 \hline
Smoking & No & 2625 (77) & 7337 (82) \\
 & Yes & 784 (23) & 1612 (18) \\
\hline
Periconceptional Drinking & No & 2098 (62) & 5601 (63) \\
 & Yes & 1303 (38) & 3316 (37) \\
\hline
Folic Acid Supplement Use & No & 2320 (68) & 5878 (66) \\
 & Yes & 1089 (32) & 3071 (34) \\
 \hline
 Subpopulation & Arkansas & 388 (11) & 1163 (13) \\
 	& California & 530 (16) & 1031 (12) \\
	& Iowa & 344 (10) & 1019 (11) \\
	& Massachusetts & 443 (13) & 1082 (12) \\
	& New Jersey & 163 (5) & 510 (6) \\
	& New York  & 272 (8) & 776 (9) \\
	& Texas & 367 (11) & 1061 (12) \\
	& Georgia & 390 (11) & 945 (11) \\
	& North Carolina & 241 (7) & 721 (8) \\
	& Utah & 271 (8) & 641 (7) \\
 \endline
\end{tabular}
\caption{Descriptive statistics of NBDPS participants included in study}
\label{tab:demog}
\end{table}

Dietary intake was reported using the Willet food frequency questionnaire. Each food item listed in the questionnaire contained 16 responses on frequency consumption ranging from `never or less than one month' to `six or more times a day'. A separate questionnaire was used to assess level of intake of cereal, soft drinks, coffee and tea. A total of 63 food items were used for analysis. Each food item was categorized into four ordinal categories based on relative daily consumption: no consumption (or very low levels rounding to 0) and tertiles of nonzero consumption. Consumption percentiles were calculated to control for differences across energy intake. Percentiles were calculated by dividing the grams per day of an individual food item by the total grams per day of food consumed.  Grams per day of a food item consumed were calculated by multiplying the standard portion size listed on the questionnaire and the reported frequency of consumption \citep{sotres2013maternal}. Posterior results were computed after an MCMC run of 20,000 iterations and a 5,000 burn-in. 

\subsection{Comparative Analysis}
\begin{figure}[H]
  \centering
  \includegraphics[width=\textwidth,height=0.5\textheight]{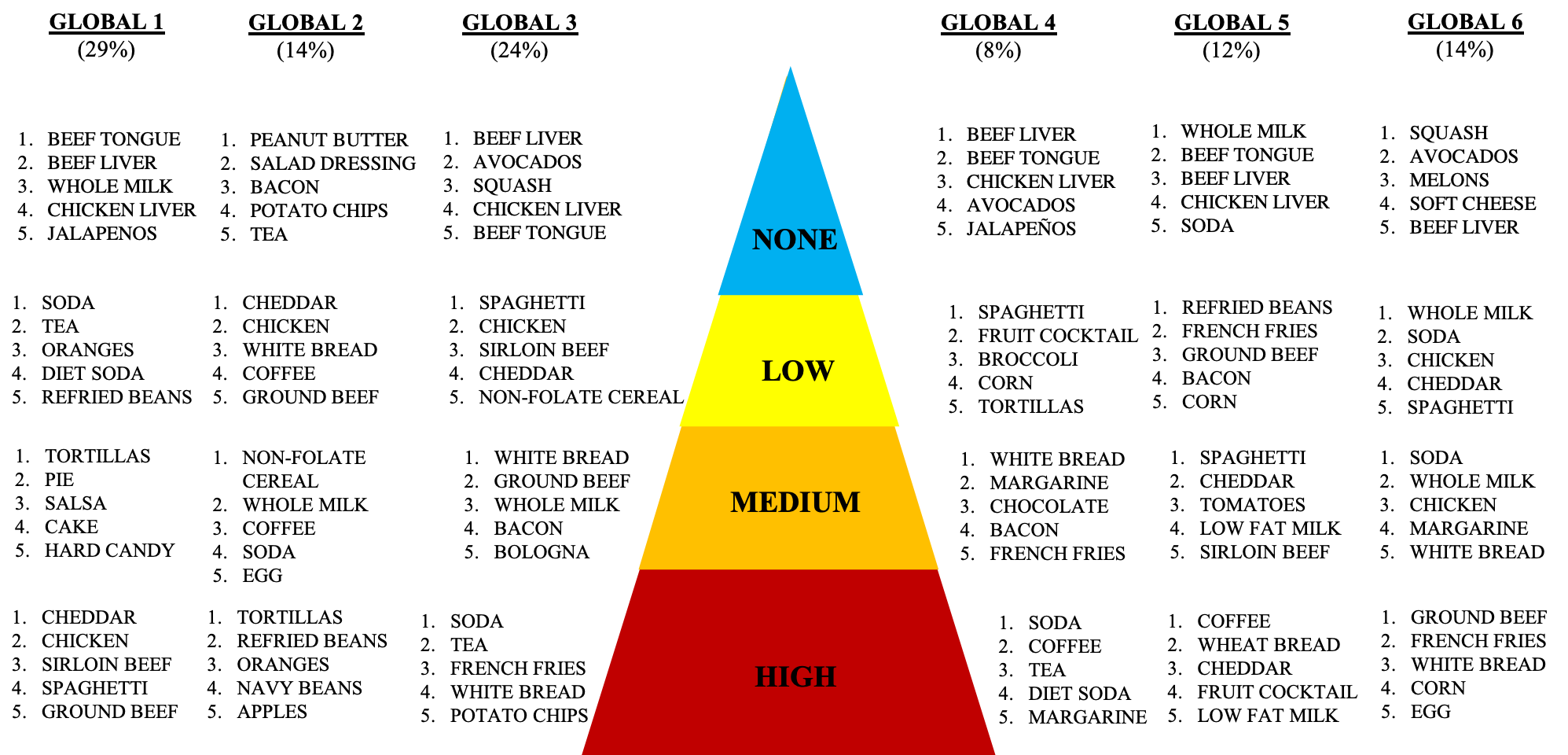}
   \caption{Top 5 foods based on likelihood of consumption at the specified level of global profiles derived from sRPC predictor model. \label{fig:pyramid}}
\end{figure}

Six global profiles were deemed appropriate in the supervised RPC joint model. Using the global cluster assignments from the MCMC output, subjects were assigned to one of the six dietary profiles via hierarchical clustering using the distance criterion described in \ref{parmest}. Figure \ref{fig:pyramid} provides an illustration of the top five foods most likely to be consumed at each consumption level for each global profile along with membership probabilities. 

Global profile 1, which had the largest proportion of participants assigned, had a high consumption of land meat (chicken, ground beef, sirloin beef), and a medium consumption of sweets (pie, cake, hard candy). Global profile 2 favored foods typically found in a Hispanic/Latino diet (high consumption of tortillas, refried beans, navy beans). Global profile 3 favored foods typically found in a fast-food diet with a high consumption of starches (white bread, french fries, potato chips), caffeine (soda, tea), and a medium consumption of processed red meat (bacon, bologna, ground beef). Global profile 4, which had the smallest proportion of participants assigned, resembled a high consumption of caffeine (soda, coffee, tea) and a medium consumption of white bread, margarine, and chocolate. Global profile 5 had the most prudent of the profiles, with a high consumption of wheat bread, fruit cocktail, and low fat milk. Global profile 6, like global profile 3, also shared an American fast-food style diet with high consumption of ground beef, french fries, white bread corn, and eggs, but only a medium consumption of soda and whole milk.

Within each subpopulation, we found varying distributions of global profile assignment. Global profile 1 was the most prominent profile represented in seven of the ten states, ranging from 27\% (New Jersey) to 48\% (Utah). Global profile 2, which favored Hispanic/Latino frequently consumed foods was the most prominent profile in California. Arkansas and Texas favored global profile 3 with the medium consumption of processed red meats and least likely to be in the most prudent global profile 5 (Arkansas=6\%; Texas7\%). Global profile 5 was the second most prominent profile in Massachusetts (18\%) and Utah (16\%). 

\begin{figure}[H]
 \centering
 \includegraphics[width=\textwidth,height=0.5\textheight]{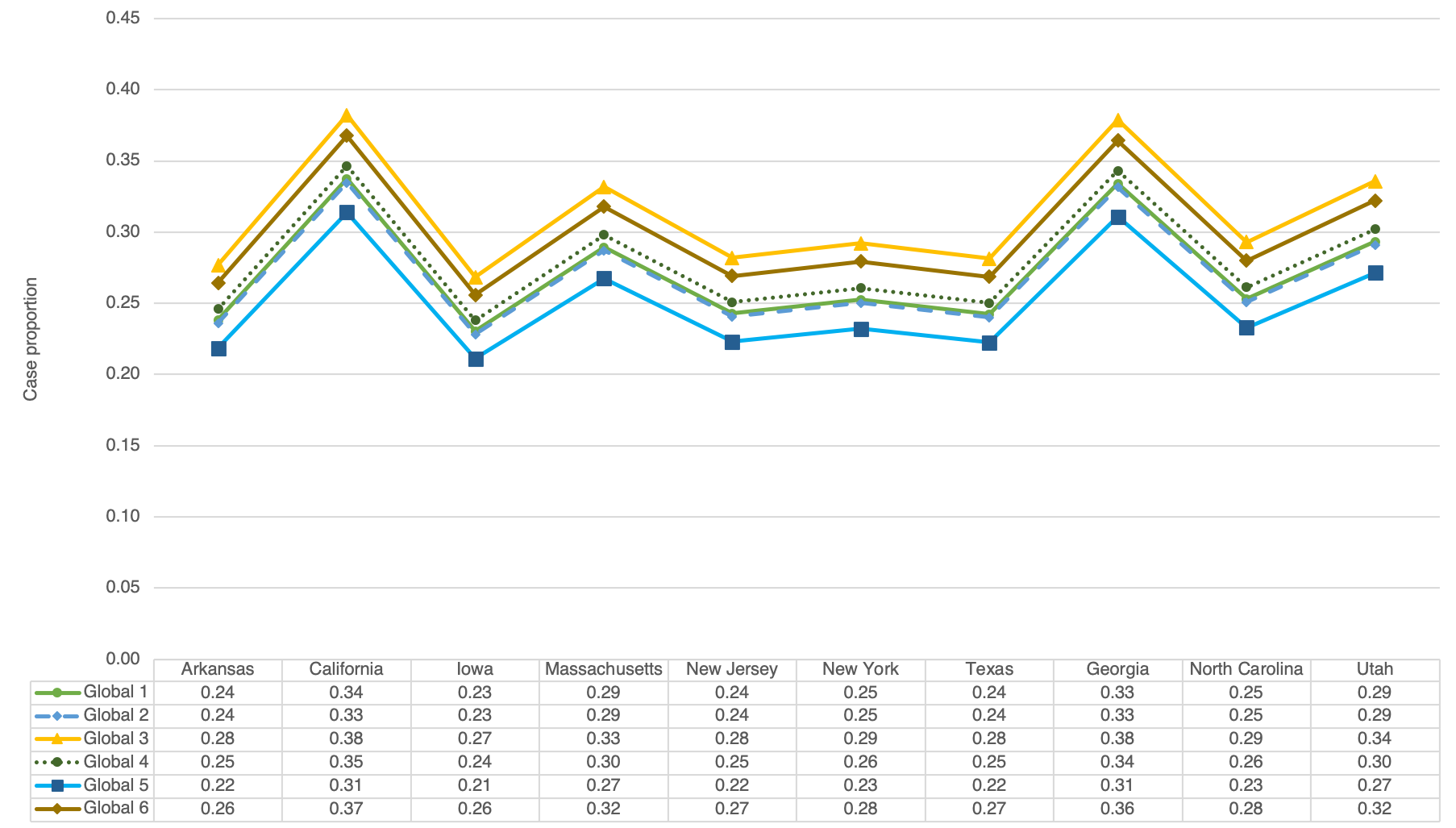}
  \caption{Plot of orofacial cleft by state based on the following demographic: Non-Hispanic White, less than 25 years old, at least high school education, non-smoking \label{fig:probitline}}
\end{figure}

Several demographics have been found to show an association with birth defects. Descriptive statistics of the demographic data are provided in Table \ref{tab:demog}. We provide the posterior estimates of the sRPC probit regression model parameters, under a reference cell coding scheme, are provided in Table \ref{tab:probdems} provides the posterior estimates. We select the following  demographic as the referent group: Massachusetts residents, with at least a high school education, non-smoker, non-Hispanic White, aged less than 25 years old, assigned to global profile 5. We illustrate the trends of the proportion with a defect for subpopulation and global profile assignment, by holding constant the demographics most largely represented in the study population (Non-smoking, Non-Hispanic White, at least high school education) in Figure \ref{fig:probitline}.

\begin{table}[H]
\centering
\renewcommand{\arraystretch}{0.65}
 \begin{tabular}{||l c c| c||} 
 \hline
 Coefficient & Median & 95\% Credible Interval & $P(\xi > 0)$ \\ [0.5ex] 
 \hline\hline
 Intercept & -0.62 & (-3.02, 0.29) & 0.12 \\
Arkansas &  -0.16 & (-0.46, 0.00) & 0.02 \\ 
California & 0.14 & (0.00, 0.39) & 0.97\\
Iowa & -0.18 & (-0.51, -0.01) & 0.01\\
New Jersey & -0.14 & (-0.43, 0.01) & 0.04\\
New York & -0.11 & (-0.33, 0.01) & 0.05\\
Texas & -0.14 & (-0.43, 0.00) & 0.03\\
Georgia & 0.13 & (-0.01, 0.40) & 0.97 \\
North Carolina & -0.11 & (-0.33, 0.02) & 0.05\\
Utah & 0.01 & (-0.09, 0.12) & 0.59 \\
 \hline
Less than High School education & 0.12& (0.01, 0.34) & 0.99\\
Smoker & 0.15 & (0.02, 0.40) & 0.99 \\
Age $\ge 25$ & 0.03 & (-0.03, 0.11) & 0.81\\
Non-Hispanic Black &-0.54 & (-1.62, -0.10) & 0.00\\
Hispanic & -0.06 & (-0.25, 0.05) & 0.14\\
Other Race & -0.03 & (-0.15, 0.07) & 0.29 \\
\hline
supRPC Global 1 & 0.07 & (-1.15, 1.85) & 0.64\\
supRPC Global 2 & 0.06 & (-1.08, 1.68) & 0.62 \\
supRPC Global 3 & 0.18 & (-0.83, 2.06) & 0.82\\
supRPC Global 4 & 0.09 & (-1.20, 1.82) & 0.67\\
supRPC Global 6 & 0.15 & (-1.14, 1.87) & 0.78\\
\endline
\end{tabular}%
\caption{Parameter estimates of supervised RPC joint model, adjusting for demographic confounders. Referent group: Massachusetts, at least high school education, non-smoker, non-Hispanic White, age $<$ 25, global profile 5 assignment. }
\label{tab:probdems}
\end{table}

Georgia and California had the highest proportions of orofacial cleft cases, and Iowa had the lowest. The profiles favoring American-style fast food items (Global 3 and 6) had the highest proportin of orofacial cleft cases. The most prudent profile (Global 5) had the lowers proportion of orofacial cleft cases. This is consistent with findings in \citet{chen2014exploring, chen2016maternal}, where higher consumptions of  caffeine, red meat were found to have adverse reproductive outcomes, and higher intake of fruits, vegetables, and whole grain had more positive reproductive outcomes. 

We now consider the results of the standard LCA joint model. {\it{Post hoc}} testing indicated 4 global clusters was the best fit. The effects of these clusters with the demographic characteristics are summarized in Table \ref{tab:lcaparm}.  All four clusters favored a non-consumption of 19 foods (butter, soft cheese, whole milk, margarine, chicken liver, frankfurter, melons, cabbage, cooked carrots, peas, spinach, yams, squash, beef liver, beef tongue, coffee, tea, pie, diet soda, fortified cereal). Latent cluster (LC) 4 was the largest cluster with 32\% of mothers assigned. This cluster was the most prudent with a high consumption favored in wheat bread, spaghetti, fruit cocktail, and carrots. LC 2 was the smallest cluster with 15\% of mothers assigned and favored foods typically found in a Hispanic/Latino diet (tortillas, refried beans, salsa, navy beans). LC 1 and LC 3 had higher proportion of orofacial cleft cases, with LC 3 having the higher of the two. These two clusters shared a modal high consumption of ground beef and french fries. LC 2 and LC 4 have nearly identical effects on the proportion of orofacial cleft cases. Both of these clusters favored a low consumption of ground beef and starches (potatoes, french fries, potato chips).

The patterns reflected in the LCA model showed some similarities in consumption modes and member assignment to those reflected in the sRPC model. The same modal consumption level of 45 (71\%) foods were shared amongst LC1 and sRPC global 3. LC 2 shared the same mode with sRPC global 2 for 43 (68\%) foods. LC 3 shared the same mode with sRPC global 6 with 52 (83\%) foods. LC 4 shared the same mode with sRPC global 1 with 30 (48\%) foods. As a result of these consumption mode mappings, a majority of those LC-assigned participants shared the same sRPC global profile assignment: 56\% of LC-1 assigned participants belonged to sRPC global 3, 62\% of LC-2 assigned participants belonged to sRPC global 2, 48\% of LC-3 assigned participants belonged to sRPC global 6, 58\% of LC-4 assigned participants belonged to sRPC global 1. 

Both LC 4 and sRPC global profile 1 had the largest representation of participants (32\% and 29\% respectively). However, the foods favored to be consumed at a high consumption level are very different. The five foods most likely to be consumed at a high consumption level for LC 4 are cheddar, wheat bread, spaghetti, fruit cocktail and raw carrots. These foods appear to be more consistent with sRPC's prudent global profile 5, compared to the high consumption of beef, chicken indicated in global profile 1. However, this prudent LC pattern has a reduced strength in posterior probability of foods consumed at that level. For example, the probabilities of those 5 foods being consumed at the high consumption level for LC 4 ranged from 39-44\%. The probabilities of the top five foods consumed at the high cosumption level for sRPC global 1 range from 60-73\%. This is a consequence of the LCA model's limitation of global clustering. With a large representation of different state-specific dietary information included in a single cluster, the true pattern is diluted with other competing subpopulation influences. This makes it more challenging to identify where and how the true pattern effects differ as it relates to orofacial cleft case representation. All other demographic characteristics showed similar results to the supervised RPC model. Mothers in California and Georgia had a higher proportion of orofacial cleft cases compared to the other states. Similarly, mothers who had less than a high school education, smoked, and identified as Non-Hispanic white had significantly higher proportion of orofacial cleft. 

\begin{table}[H]
\centering
\renewcommand{\arraystretch}{0.5}
 \begin{tabular}{||l c c| c||} 
 \hline
 Coefficient & Median & 95\% Credible Interval & $Pr(\xi > 0)$ \\ [0.5ex] 
 \hline\hline
 Intercept & -0.62 & (-2.27, 0.23) & 0.11\\
Arkansas & -0.19 & (-0.59, -0.02) & 0.01 \\ 
California & 0.13 & (-0.01, 0.33) & 0.97 \\
Iowa & -0.19 & (-0.57, -0.02) & 0.01\\
New Jersey & -0.15 & (-0.47, 0.01) & 0.04\\
New York & -0.12 & (-0.36, 0.01) & 0.04 \\
Texas & -0.15 & (-0.45, 0.00) & 0.03\\
Georgia & 0.12 & (-0.01, 0.32) & 0.96\\
North Carolina & -0.11 & (-0.36, 0.02) & 0.05\\
Utah & 0.03 & (-0.06, 0.15) &0.75\\
 \hline
Less than High School education & 0.12& (0.01, 0.32) & 0.99\\
Smoker & 0.14 & (0.02, 0.36) & 0.99 \\
Age $\ge 25$ & 0.01 & (-0.03, 0.08) & 0.72 \\
Non-Hispanic Black & -0.50 & (-1.56, -0.08) & 0.00 \\
Hispanic & -0.04 & (-0.14, 0.04) & 0.18\\
Other Race & -0.04 & (-0.17, 0.04) & 0.17\\
 \hline
LC 1 & 0.11 & (-0.02, 0.46) & 0.95\\
LC 2 & 0.01 & (-0.12, 0.21) & 0.60\\
LC 3 & 0.17 & (0.00, 0.65) & 0.98\\
\endline
\end{tabular}
\caption{Coefficient parameter estimates of supervised LCA Model, where LC denotes latent class. Referent group: Iowa, at least high school education, non-smoker, non-Hispanic White, folic acid use, less than 25 years of age, Latent class 4 assignment. }
\label{tab:lcaparm}
\end{table}

Inconsistencies were identified with the information criterion applied to NBDPS data, due to the increased dimensionality and reparameterization of the data structure \citep{kim2019deviance}. Average extended Bayesian information criterion (AEBIC) strongly favored the supervised RPC model by a factor of 10 \citep{xue2017average}. Deviance information criterion favored the LCA model by a factor of 1.4. As an alternative, model fitness was evaluated using a posterior predictive check comparing the deviance of the training set (90\% of the data) with the deviance of the testing set (10\% of remaining data) using the posterior median estimates derived from the training set. This process was performed on 100 different permutations of the data. A visualization comparing the two models is provided in the Web Figure 3. The mean difference of deviance for the supervised RPC ranged from (-1.87, 0.78) with a standard deviation of 0.58, whereas the supervised LCA ranged (-3.58, 2.75) with a standard deviation of 1.50. This difference in dispersion between the models illustrates the improved fit to the data achieved using the supervised RPC compared to the supervised LCA. Posterior predictive check for the outcome of interest was also performed and compared which showed a slight improvements under the RPC model compared to the LCA (0.60 vs 0.61).  

\subsection{Local Profile Patterns}
\label{ss:local}

From an exploratory perspective, we can get a more holistic picture of the dietary profiles for each subpopulation and understand what foods may or may not be contributing to an association of an orofacial cleft. A subset of foods was identified as having a tendency to deviate from a global profile in favor of a separate consumption pattern at the subpopulation level. Within each subpopulation, one local profile was identified. The complete set of local profile patterns, identified by modal consumption level, is provided for reference in Web Figure 4. We focus here on those foods that had a deviation tendency of at least 60\% probability of allocation to the subpopulation/local level (Figure \ref{fig:devs}). Only foods with varying consumption patterns across subpopulations are included in Figure \ref{fig:devs}. At this threshold, four food items favored a local tendency and non-consumption across all subpopulations: chicken liver, beef liver, diet soda and fortified cereal. 

\begin{figure}[H]
\begin{center}
\includegraphics[width=0.55\textwidth]{LocDev60_v3}
\caption{Heat map illustrating foods with a tendency $(\nu_j^{(s)}<0.40)$ to deviate from supRPC global profiles by subpopulation. Legend: Blue-no consumption, Orange-medium consumption, Red-high consumption. Colorless spaces indicate foods not allocated to the local (subpopulation) level.
\label{fig:devs}}
\end{center}
\end{figure}

Mothers from Utah, which were most likely to be assigned to global profile 1 (48\%), had the largest number of foods (22 foods) to prefer an allocation to the local level, and assume the pattern reflected in their local profile. At the local level, these mothers favored a high consumption of eggs, dairy (cheese, yogurt, low fat milk), bananas, chocolate, and non-folate cereal. Mothers from Georgia, which also were more likely to be assigned to global profile 1 (30\%), had the largest number of foods (55 foods) to allocate to the global level, and assume the pattern reflected in their assigned global profile. 

Georgia and California which had the highest proportion of an orofacial cleft shared similar consumption modes for two foods at the local level: a nonconsumption of soft cheese and a high consumption of non-folate cereal. Iowa which had the lowest proportion of orofacial cleft cases favored a notably high consumption of potatoes, sirloin beef, and low fat milk. Other notably high modes of consumption from other states include refried beans and tortillas in Texas, tortillas in California, spaghetti and salad dressing in Massachusetts. Illustrations and tables detailing the complete consumption  patterns of the global and local profiles are provided in Web Figures 3 and 4. 

\section{Discussion}
\label{s:discuss}

The supervised RPC model provides a bridge connecting clustering analysis with confirmatory regression analysis. The multilevel approach allows a separation of exposures or behaviors shared across an overall population from those that uniquely deviate within a respective subpopulation. The joint model is able to create a clustering system that reflects population patterns associated with a health outcome, as demonstrated here with an orofacial cleft defect. 

Applying this to the NBDPS maternal dietary information, we learn that diets that favor a high consumption of fruits and vegetables paired with a low consumption of land meats (chicken, beef) were associated with lower proportion of orofacial defects. These results are consistent with prior NBDPS studies that also showed a prudent diet or high diet quality score was related to lower proportion of orofacial cleft defects \citep{carmichael2012reduced,collier2009maternal}. The results from the supervised RPC model add to the previous studies by highlighting consumption patterns of a subset of foods that are commonly consumed across the United States.  

Contrary to unsupervised models, the supervised RPC model is sensitive to both the population and outcome of interest. A different health outcome would reflect a potentially different clustering model than the one described in this paper, as clustering would account for association with a different outcome. Future directions focus on robustness of clustering methods more generally and methods allowing varying coefficients to accommodate local deviations.


\backmatter


\section*{Acknowledgements}

The authors gratefully acknowledge this work was supported in part through cooperative agreements from the Centers for Disease Control and Prevention to the centers participating in the National Birth Defects Prevention Study, and by the National Institute of Environmental Health Sciences (R01ES020619). Dietary nutrition information was made possible by University of North Carolina Epidemiology Core (Grant No: DK56350;  NIEHS T32ES007018). We thank Leo Duan for assistance in data replication. Finally, we thank Louise Ryan and David Dunson for helpful comments on earlier drafts of this work. \vspace*{-8pt}

\section*{Data Availability Statement}
Data used to support and replicate findings from the simulation study are made publicly available on GitHub repository \url{https://github.com/bjks10/supRPC}. Privacy restrictions apply to the availability of NBDPS data, which were used under a data use agreement for this study. Data are available with permission from the National Birth Defects Prevention Study.


\bibliographystyle{biom} 
\bibliography{sRPCrefs}




\section*{Supporting Information}
Additional supporting information may be found online in the Supporting Information section at the end of the article.

\label{lastpage}

\end{document}


\appendix

\section*{Supporting Information for Supervised Robust Profile Clustering by Stephenson, Herring, and Olshan}

\subsection{Web Figure 1}
\begin{figure}[H]
\includegraphics[width=0.95\textwidth]{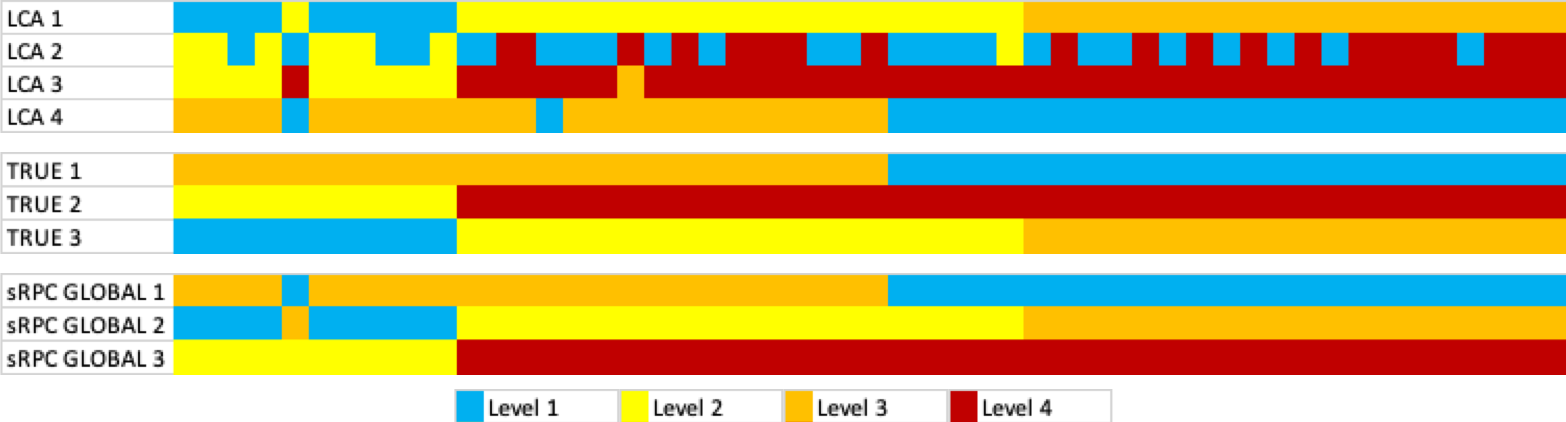}
\caption{Heatmap of expected cluster pattern modes (middle) compared to cluster patterns derived in the LCA model (top) and supervised RPC model (bottom). Note: this is a single replicate randomly selected from the 500 simulated replicates analzyed.\label{fig:mode}}
\end{figure}

\subsection{Web Figure 2}
\begin{figure}[H]

  \begin{subfigure}[b]{0.5\textwidth}
    \includegraphics[width=\textwidth,height=0.7\textheight]{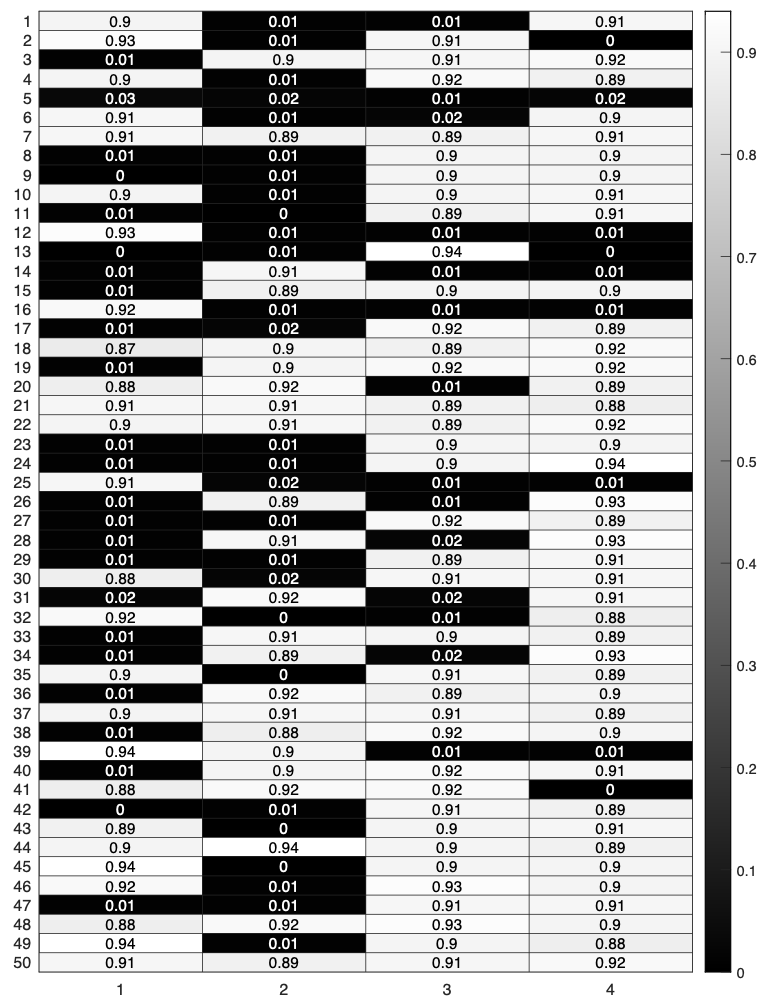}
    \caption{Predicted}
    \label{fig:1}
  \end{subfigure}
  \begin{subfigure}[b]{0.5\textwidth}
    \includegraphics[width=\textwidth,height=0.7\textheight]{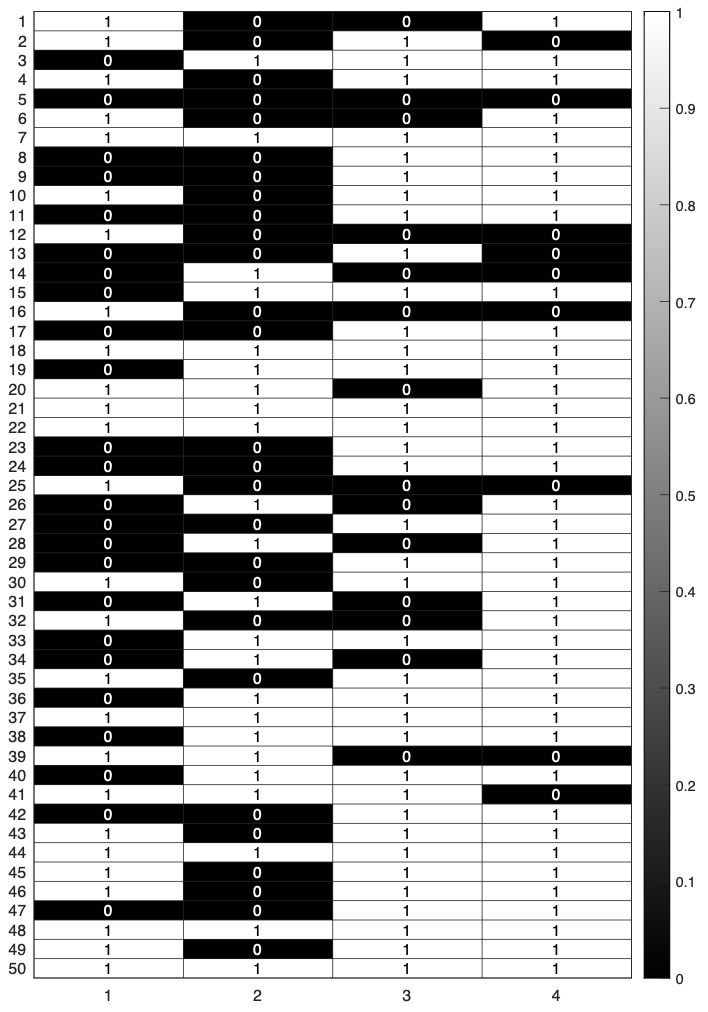}
    \caption{Expected}
    \label{fig:2}
  \end{subfigure}
  \caption{Heatmap example of a simulated dataset comparing probability of variable allocated to a global set (white, $G_{ij}=1$) or local set (black, $G_{ij}=0$). Subfigure (a) on the left illustrates the predicted probability derived from the supervised RPC model. Subfigure (b) on the right illustrates the true allocations.\label{fig:nu}}
\end{figure}

\subsection{Web Figure 3}
\begin{figure}[H]
	\includegraphics[height=0.85\textheight]{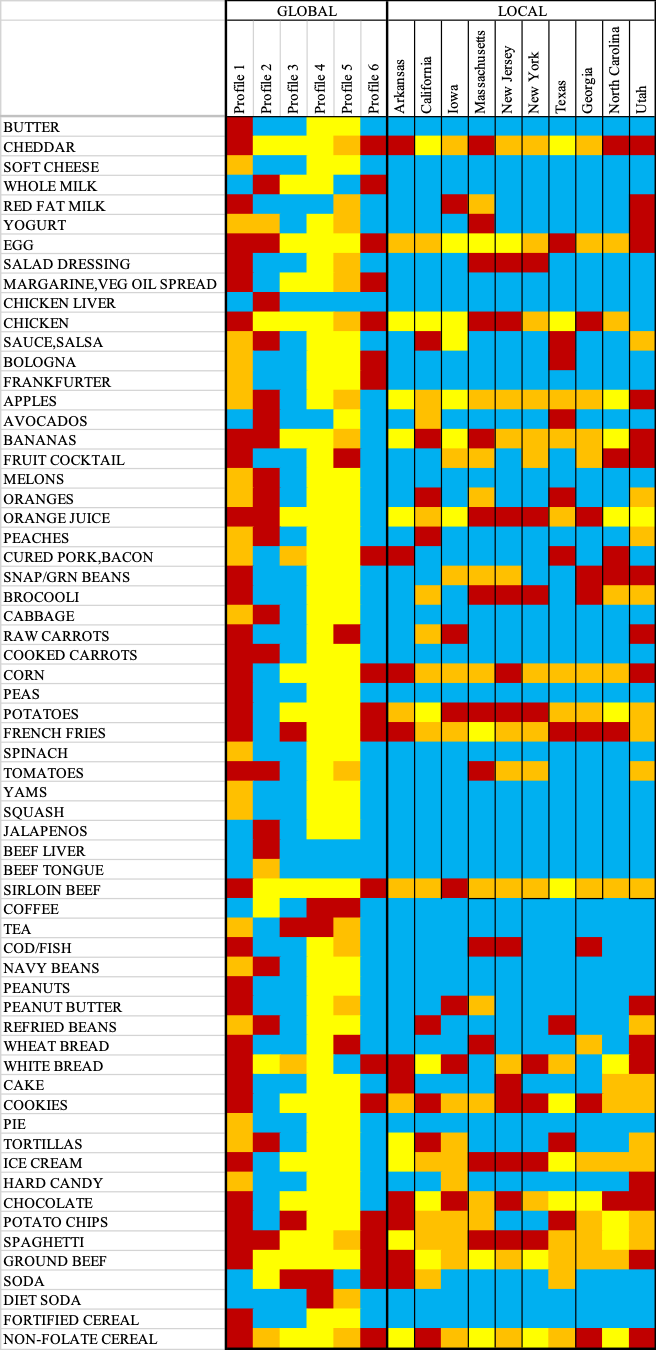}
	\caption{Modal consumption patterns of all global and local profiles derived from sRPC for 10 states included in study. Legend: Blue-no consumption, Orange-medium consumption, Red-high consumption.}
	\label{fig:devs}
\end{figure}

\subsection{Web Figure 4}

\begin{figure}[H]
	\includegraphics[height=0.85\textheight]{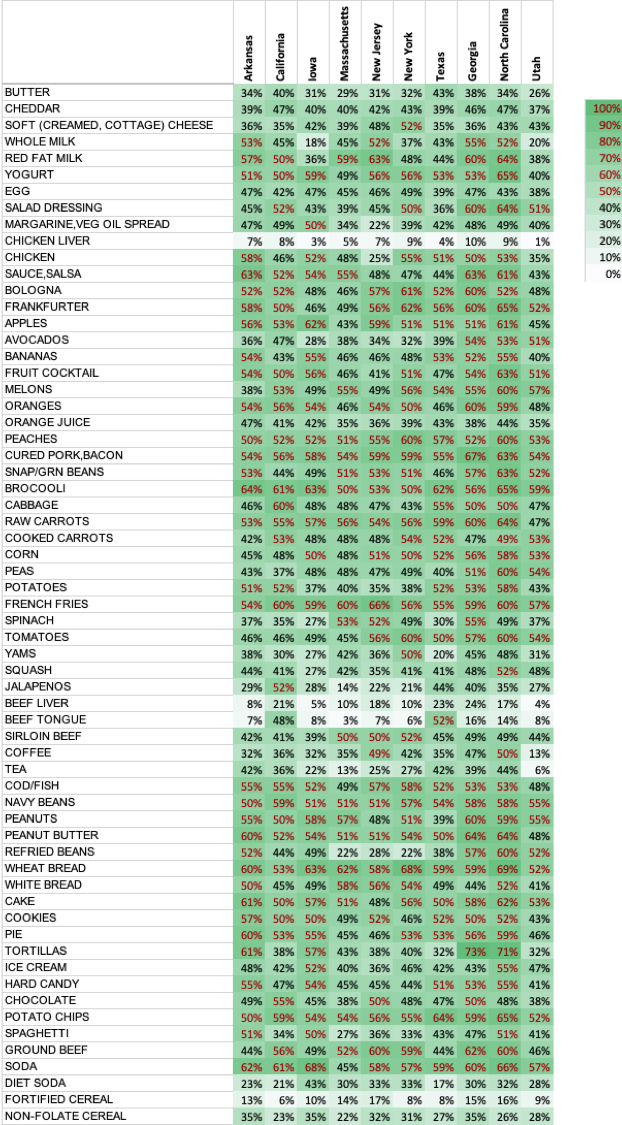}
	\caption{Heatmap illustrating sRPC probability parameter $\nu_j^{(s)}$. Rows=food items, columns=subpopulations. Higher probabilities indicate a higher probability of the food assuming pattern at the global level. Lower probabilities indicate higher probability of food assuming pattern at the local (subpopulation) level}
	\label{fig:nu}
	\end{figure}
\subsection{Web Figure 5}

\begin{figure}[H]
	\includegraphics[width=0.95\textwidth]{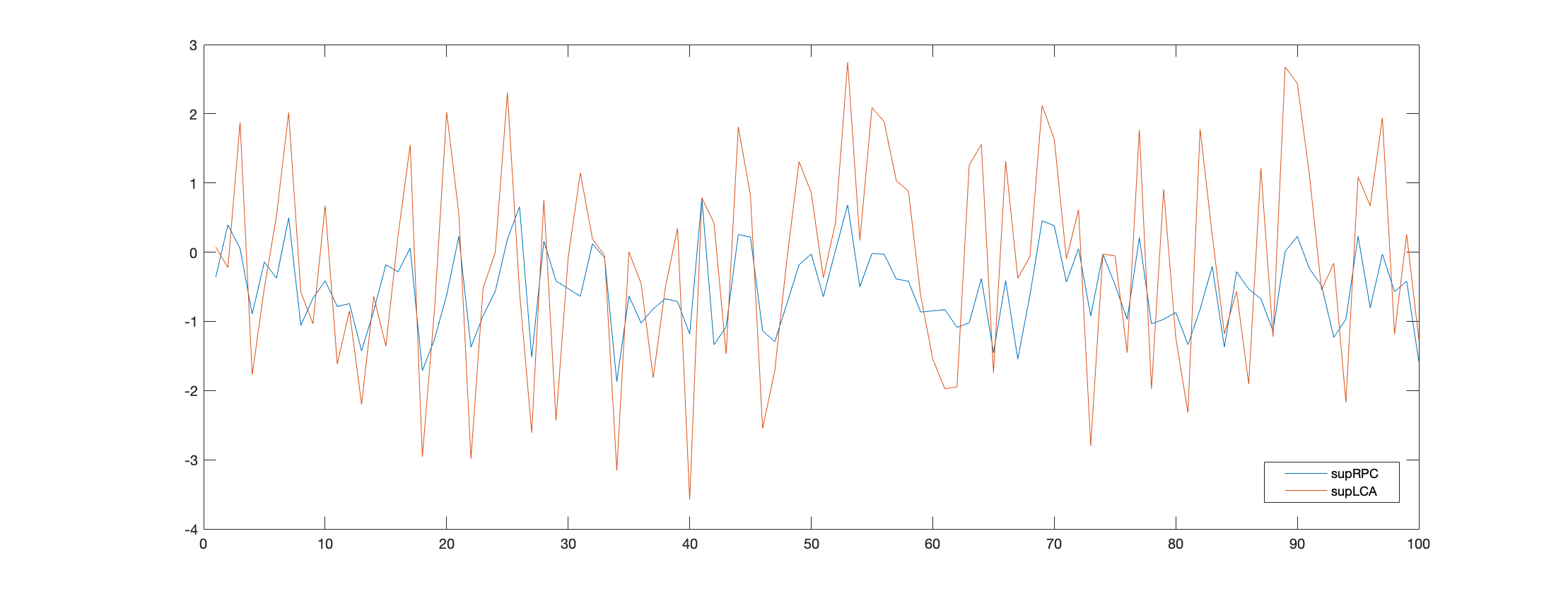}
	\caption{Validation results based on 100 test and training sets of NBDPS data. Values plotted are the difference in deviance between the testing and training data for the respective set. Supervised RPC model results are illustrated in blue. Supervised LCA model results are illustrated in orange.}
	\label{fig:devpl}
\end{figure}
\subsection{Web Table 1}

\begin{table}[htbp]
  \centering
\scalebox{0.7}{%
\begin{adjustbox}{angle=90}
    \begin{tabular}{l|rr|rr|rr|rr|rr|rr|r|rr|rr|rr|rr|}
\cmidrule{2-13}\cmidrule{15-22}          & \multicolumn{12}{c|}{Robust Profile Clustering}                                               &       & \multicolumn{8}{c|}{Latent Cluster Analysis} \\
\cmidrule{2-22}          & \multicolumn{2}{c|}{Global 1} & \multicolumn{2}{c|}{Global 2} & \multicolumn{2}{c|}{Global 3} & \multicolumn{2}{c|}{Global 4} & \multicolumn{2}{c|}{Global 5} & \multicolumn{2}{c|}{Global 6} &       & \multicolumn{2}{c|}{LC 1} & \multicolumn{2}{c|}{LC 2} & \multicolumn{2}{c|}{LC 3} & \multicolumn{2}{c|}{LC 4} \\
\cmidrule{2-22}    Arkansas & 364   & 23\%  & 118   & 8\%   & \textbf{571} & \textbf{37\%} & 170   & 11\%  & 100   & 6\%   & 228   & 15\%  &       & \textbf{728} & \textbf{47\%} & 99    & 6\%   & 432   & 28\%  & 292   & 19\% \\
    California & 340   & 22\%  & \textbf{485} & \textbf{31\%} & 310   & 20\%  & 117   & 7\%   & 144   & 9\%   & 165   & 11\%  &       & 312   & 20\%  & \textbf{647} & \textbf{41\%} & 318   & 20\%  & 284   & 18\% \\
    Iowa  & \textbf{426} & \textbf{31\%} & 78    & 6\%   & 404   & 30\%  & 145   & 11\%  & 163   & 12\%  & 147   & 11\%  &       & \textbf{567} & \textbf{42\%} & 51    & 4\%   & 272   & 20\%  & 473   & 35\% \\
    Massachusetts & \textbf{512} & \textbf{34\%} & 182   & 12\%  & 264   & 17\%  & 100   & 7\%   & 275   & 18\%  & 192   & 13\%  &       & 294   & 19\%  & 38    & 2\%   & 399   & 26\%  & \textbf{794} & \textbf{52\%} \\
    New Jersey & \textbf{180} & \textbf{27\%} & 126   & 19\%  & 124   & 18\%  & 44    & 7\%   & 83    & 12\%  & 116   & 17\%  &       & 138   & 21\%  & 45    & 7\%   & \textbf{273} & \textbf{41\%} & 217   & 32\% \\
    New York & \textbf{312} & \textbf{30\%} & 146   & 14\%  & 217   & 21\%  & 82    & 8\%   & 126   & 12\%  & 165   & 16\%  &       & 272   & 26\%  & 52    & 5\%   & 321   & 31\%  & \textbf{403} & \textbf{38\%} \\
    Texas & 272   & 19\%  & 345   & 24\%  & \textbf{395} & \textbf{28\%} & 99    & 7\%   & 93    & 7\%   & 224   & 16\%  &       & 291   & 20\%  & \textbf{588} & \textbf{41\%} & 395   & 28\%  & 154   & 11\% \\
    Georgia & \textbf{407} & \textbf{30\%} & 248   & 19\%  & 243   & 18\%  & 66    & 5\%   & 161   & 12\%  & 210   & 16\%  &       & 282   & 21\%  & 214   & 16\%  & 367   & 27\%  & \textbf{472} & \textbf{35\%} \\
    North Carolina & \textbf{292} & \textbf{30\%} & 151   & 16\%  & 228   & 24\%  & 66    & 7\%   & 116   & 12\%  & 109   & 11\%  &       & 289   & 30\%  & 134   & 14\%  & 212   & 22\%  & \textbf{327} & \textbf{34\%} \\
    Utah  & \textbf{436} & \textbf{48\%} & 85    & 9\%   & 119   & 13\%  & 58    & 6\%   & 146   & 16\%  & 68    & 7\%   &       & 152   & 17\%  & 46    & 5\%   & 146   & 16\%  & \textbf{568} & \textbf{62\%} \\
    \midrule
    No HS edu & 846   & 16\%  & 1325  & 25\%  & \textbf{1570} & \textbf{30\%} & 394   & 8\%   & 245   & 5\%   & 868   & 17\%  &       & 1566  & 30\%  & 1524  & 29\%  & \textbf{1580} & \textbf{30\%} & 578   & 11\% \\
    HSEdu & \textbf{2695} & \textbf{38\%} & 639   & 9\%   & 1305  & 18\%  & 553   & 8\%   & 1162  & 16\%  & 756   & 11\%  &       & 1759  & 25\%  & 390   & 5\%   & 1555  & 22\%  & \textbf{3406} & \textbf{48\%} \\
    \midrule
    Non-smoker & \textbf{3113} & \textbf{31\%} & 1818  & 18\%  & 1859  & 19\%  & 640   & 6\%   & 1240  & 12\%  & 1292  & 13\%  &       & 2115  & 21\%  & 1789  & 18\%  & 2510  & 25\%  & \textbf{3548} & \textbf{36\%} \\
    Smoker & 428   & 18\%  & 146   & 6\%   & \textbf{1016} & \textbf{42\%} & 307   & 13\%  & 167   & 7\%   & 332   & 14\%  &       & \textbf{1210} & \textbf{51\%} & 125   & 5\%   & 625   & 26\%  & 436   & 18\% \\
    \midrule
    Underweight & 151   & 23\%  & 106   & 16\%  & \textbf{205} & \textbf{31\%} & 43    & 6\%   & 53    & 8\%   & 111   & 17\%  &       & \textbf{203} & \textbf{30\%} & 82    & 12\%  & 227   & 34\%  & 157   & 23\% \\
    Normal & \textbf{1990} & \textbf{31\%} & 822   & 13\%  & 1448  & 23\%  & 441   & 7\%   & 805   & 13\%  & 847   & 13\%  &       & 1625  & 26\%  & 714   & 11\%  & 1603  & 25\%  & \textbf{2411} & \textbf{38\%} \\
    Obese & \textbf{1367} & \textbf{29\%} & 669   & 14\%  & 1182  & 25\%  & 448   & 9\%   & 524   & 11\%  & 606   & 13\%  &       & \textbf{1470} & \textbf{31\%} & 738   & 15\%  & 1193  & 25\%  & 1395  & 29\% \\
    \midrule
    No Drinking & \textbf{2129} & \textbf{28\%} & 1531  & 20\%  & 1757  & 23\%  & 469   & 6\%   & 675   & 9\%   & 1138  & 15\%  &       & 1861  & 24\%  & 1508  & 20\%  & \textbf{2183} & \textbf{28\%} & 2147  & 28\% \\
    Periconceptional Drinking & \textbf{1395} & \textbf{30\%} & 428   & 9\%   & 1109  & 24\%  & 475   & 10\%  & 731   & 16\%  & 481   & 10\%  &       & 1454  & 31\%  & 400   & 9\%   & 941   & 20\%  & \textbf{1824} & \textbf{39\%} \\
    \midrule
    White & \textbf{2533} & \textbf{35\%} & 350   & 5\%   & 1848  & 25\%  & 676   & 9\%   & 1098  & 15\%  & 802   & 11\%  &       & 2468  & 34\%  & 82    & 1\%   & 1563  & 21\%  & \textbf{3194} & \textbf{44\%} \\
    Black & 322   & 27\%  & 82    & 7\%   & \textbf{342} & \textbf{29\%} & 70    & 6\%   & 48    & 4\%   & 324   & 27\%  &       & 367   & 31\%  & 21    & 2\%   & \textbf{547} & \textbf{46\%} & 253   & 21\% \\
    Hispanic & 447   & 15\%  & \textbf{1342} & \textbf{45\%} & 497   & 17\%  & 140   & 5\%   & 175   & 6\%   & 359   & 12\%  &       & 289   & 10\%  & \textbf{1714} & \textbf{58\%} & 712   & 24\%  & 245   & 8\% \\
    Other & \textbf{239} & \textbf{26\%} & 190   & 21\%  & 188   & 21\%  & 61    & 7\%   & 86    & 10\%  & 139   & 15\%  &       & 201   & 22\%  & 97    & 11\%  & \textbf{313} & \textbf{35\%} & 292   & 32\% \\
    \midrule
    Non FA use & 1933  & 24\%  & 1572  & 19\%  & \textbf{2206} & \textbf{27\%} & 651   & 8\%   & 612   & 7\%   & 1224  & 15\%  &       & \textbf{2411} & \textbf{29\%} & 1669  & 20\%  & 2310  & 28\%  & 1808  & 22\% \\
    FA Use & \textbf{1608} & \textbf{39\%} & 392   & 9\%   & 669   & 16\%  & 296   & 7\%   & 795   & 19\%  & 400   & 10\%  &       & 914   & 22\%  & 245   & 6\%   & 825   & 20\%  & 2176  & \textbf{52\%} \\
    \midrule
    $\ge 25$ yrs age & 779   & 19\%  & 689   & 17\%  & \textbf{1428} & \textbf{35\%} & 304   & 7\%   & 169   & 4\%   & 751   & 18\%  &       & \textbf{1417}  & \textbf{34\%}  & 815  & 20\%  & 1339  & 33\%  & 549 & 13\%\\
   $< 25$ yrs age & \textbf{2762} & \textbf{34\%} & 1275  & 15\%  & 1447  & 18\%  & 643   & 8\%   & 1238  & 15\%  & 873   & 11\%  &       & 1908   & 23\%  & 1099    & 13\%   & 1796 & 22\% & \textbf{3435}   & \textbf{42}\% \\
    \end{tabular}%
\end{adjustbox}
}
  \caption{Demographic characteristics of NBDPS participants for each dietary cluster pattern of sRPC and LCA model.}
  \label{tab:demogRPCLCA}%
\end{table}%


\label{lastpage}